\documentstyle[12pt,aps,prc,epsfig,preprint]{revtex}
\tightenlines 
\begin{document}
\begin{titlepage}

\title{Cluster models of $^{~~6}_{\Lambda\Lambda}$He and $^9_{\Lambda}$Be 
hypernuclei}

\author{I. Filikhin $^{a,b}$, A. Gal $^{c}$ and V.M. Suslov $^{b}$ \\
$^a$Department of Physics, North Carolina Central University, 
Durham, NC 27707, USA \\ 
$^b$Department of Mathematical and Computational Physics, 
St. Petersburg State University, 198504 Petrodvorets, 
St. Petersburg, Russia \\ 
$^c$Racah Institute of Physics, The Hebrew University, 
Jerusalem 91904, Israel}

\maketitle

\begin{abstract}
Configuration space Faddeev calculations are performed for the 
binding energy of $^{~~6}_{\Lambda\Lambda}$He and $^9_{\Lambda}$Be 
bound states, here considered as $\alpha\Lambda\Lambda$ and 
$\alpha\alpha\Lambda$ clusters respectively, in order to study the 
dependence of the calculated binding energy on the $\alpha\Lambda$ 
potential input. For $^{~~6}_{\Lambda\Lambda}$He, using realistic 
interactions, the uncertainty in extracting the $\Lambda\Lambda$ 
$^{1}S_0$ interaction strength does not exceed 0.1 MeV, which is 
a fraction of the order of magnitude derived for other theoretical 
uncertainties. For $^9_{\Lambda}$Be, the dependence of the calculated 
binding energy on the $\alpha\Lambda$ potential is considerably larger, 
of order 1 MeV. Our results for $^9_{\Lambda}$Be suggest that the 
odd-state $\alpha\Lambda$ interaction is substantially reduced with 
respect to the even-state component. 
\newline
\newline
$PACS$: 21.80.+a, 11.80.Jy, 21.45.+v, 21.30.Fe 
\newline
\newline
Corresponding author: Avraham Gal, avragal@vms.huji.ac.il 
\newline
Tel: +972 2 6584930, Fax: +972 2 5611519
\newline
\centerline{(\today)}
\end{abstract}
\end{titlepage}

\section{Introduction}
\label{sec:int}

The recent identification of $^{~~6}_{\Lambda\Lambda}$He and the 
measurement of the binding energy of the $\Lambda\Lambda$ pair 
\cite{Tak01} has provided an experimental benchmark for 
extracting the strength of the $\Lambda\Lambda$ interaction 
within perhaps the simplest $\Lambda\Lambda$ hypernuclear system. 
A measure of the $\Lambda\Lambda$ interaction strength is given 
by the incremental binding energy: 
\begin{equation} 
\begin{array}{l} 
\Delta B_{\Lambda\Lambda}(^{~~6}_{\Lambda\Lambda}{\rm He}) 
= B_{\Lambda\Lambda}(^{~~6}_{\Lambda \Lambda}{\rm He}) - 
2B_{\Lambda}(^5_{\Lambda}{\rm He}) 
= B_{\Lambda}(^{~~6}_{\Lambda \Lambda}{\rm He}) - 
B_{\Lambda}(^5_{\Lambda}{\rm He}) 
\\
= 1.0 {\pm 0.2}^{+0.2}_{-0.1}~{\rm MeV} \;, 
\end{array} 
\label{eq:BLL} 
\end{equation} 
and is rather small compared with a similar measure of the 
$\Lambda N$ interaction strength 
\begin{equation} 
\begin{array}{l} 
\Delta B_{\Lambda N}(^5_{\Lambda}{\rm He}) 
= B_{\Lambda N}(^5_{\Lambda}{\rm He}) - 
{\bar B}_{\Lambda}(^4_{\Lambda}Z) - B_N(^4{\rm He}) 
= B_{\Lambda}(^5_{\Lambda}{\rm He}) - {\bar B}_{\Lambda}(^4_{\Lambda}Z) 
\\
= 1.73 {\pm 0.04}~{\rm MeV} \;, 
\end{array} 
\label{eq:BLN} 
\end{equation} 
where ${\bar B}_{\Lambda}$ stands for a proper spin and charge average 
over the binding energies of the $0^+$ and $1^+$ states in 
$^4_{\Lambda}$H and $^4_{\Lambda}$He \cite{FGa02b}. 
The corresponding $NN$ interaction strength is considerably larger 
than either of these values. Several subsequent calculations have 
discussed and analyzed, within three-body cluster models for the 
$A=5,6$ $\Lambda\Lambda$ hypernuclei, the model dependence of such 
extraction of the $\Lambda\Lambda$ interaction strength 
\cite{FGa02b,FGa02a,FGS03,MSA03,AG03,LY04,Yam04,FKM04}. 
In our latest report of Faddeev calculations 
for these $\Lambda\Lambda$ hypernuclei \cite{FGS03}, the 
approximation of considering only implicitly the channel coupling 
$\Lambda\Lambda-\Xi N-\Sigma\Sigma$ was estimated qualitatively, 
within the semi-realistic NSC97 OBE model, to underestimate the 
calculated binding energy by about 0.2 MeV. This has been very 
recently borne out by an explicit calculation \cite{Yam04}. 
In the present work we study the theoretical uncertainty 
resulting from the model dependence of the $\alpha\Lambda$ interaction 
which normally is constrained only by fitting to the experimental 
$B_{\Lambda}$ value of $^5_{\Lambda}$He. We will find that this 
uncertainty, for realistic potentials, amounts to about 0.1 MeV. 
In this context, we also study here the sensitivity of the 
$^9_{\Lambda}$Be binding energy calculation, within an 
$\alpha\alpha\Lambda$ model, to the assumed $\alpha\Lambda$ interaction 
since it is known that $B_{\Lambda}(^9_{\Lambda}{\rm Be})$ exhibits 
considerable sensitivity to the unknown $p$-wave component of the 
$\alpha\Lambda$ interaction \cite{HKM97}. We note several other 
few-body calculations of $^9_{\Lambda}$Be 
\cite{MBI83,PC91,OKS00,CFK02,FMK04}. 
Recently a more realistic $\alpha\Lambda$ potential consisting 
of separate $s$- and $p$-wave components \cite{MSA03} was constructed, 
inspired by the NSC97 model of the $\Lambda N$ interaction. 
The resulting $p$-wave component is weaker than the $s$-wave component. 
In the present work we use this $\alpha\Lambda$ potential for the first 
time in the $^9_{\Lambda}$Be binding energy calculation, 
in order to check the sensitivity to the $p$-wave component. 

Our methodology consists of using 
the Faddeev equations formalism in configuration space, describing 
$^{~~6}_{\Lambda\Lambda}$He and $^9_{\Lambda}$Be in terms of three-body 
systems made out of $\alpha$ clusters and $\Lambda$ particles. 
The bound-state Faddeev equations for $^9_{\Lambda}$Be, viewed as 
a three-cluster $\alpha\alpha\Lambda$ system, are discussed in Sect. 
\ref{sec:fad}, in close analogy to the discussion of the Faddeev 
equations for $^{~~6}_{\Lambda\Lambda}$He ($\alpha\Lambda\Lambda$) 
in Ref. \cite{FGS03}. Aspects of the partial-wave decomposition 
of the configuration-space Faddeev equations are relegated to 
an Appendix. In Sect. \ref{sec:pot} we specify the 
$\Lambda\Lambda$ and $\alpha\Lambda$ interaction potentials 
used in the present work to test the model dependence 
of the Faddeev calculations for binding energies. 
The results of our calculations are given in Sect. \ref{sec:res}. 
For $^{~~6}_{\Lambda\Lambda}$He, using a sufficiently expanded range of 
partial waves \cite{FGS03}, we calculated the $\Lambda\Lambda$ binding 
energy $B_{\Lambda\Lambda}$ for several $\alpha\Lambda$ potentials. 
Defining a measure $\delta B_{\Lambda\Lambda}$ 
of the $\Lambda\Lambda$ `pairing' energy, the variation of 
$\delta B_{\Lambda\Lambda}$ with the input $\alpha\Lambda$ potential 
is used to place limits on the model dependence of choosing 
$V_{\Lambda\Lambda}$ from binding energy calculations. Finally, we give 
results for $^9_{\Lambda}$Be $(\frac12)^+$ ground state and for the 
first excited ($\frac32^+,\frac52^+$) doublet of levels, for a variety 
of $\alpha\Lambda$ interaction potentials. Here the $p$-wave component, 
particularly, plays a large role, considerably larger than for 
$^{~~6}_{\Lambda\Lambda}$He. The conclusion to be drawn from the present 
Faddeev calculations of $^9_{\Lambda}$Be is that, unless three-body 
$\alpha\alpha\Lambda$ repulsive interactions contribute substantially 
(of order 1 MeV), the $p$-wave component of the $\alpha\Lambda$ 
interaction must be very weak, close to zero, with respect to the 
$s$-wave component. The paper is concluded with a summary in Sect. 
\ref{sec:sum}.

\section{Bound-state Faddeev equations for $\bf{\alpha\alpha\Lambda}$} 
\label{sec:fad} 

Bound states of three-body systems are calculated by 
solving the differential Faddeev equations \cite{MFa93}, 
which for a combination of short-range nuclear forces and Coulomb forces 
have the following form: 
\begin{equation} 
(H_0+V^s_{\gamma}(\vert\vec{x}_{\gamma}\vert)+ \sum_{\beta 
=1}^3V^c_{\beta}(\vert\vec{x}_{\beta}\vert)-E) 
\Psi_{\gamma}(\vec{x}_{\gamma},\vec{y}_{\gamma}) = 
-V^s_{\gamma}(\vert\vec{x}_{\gamma}\vert) 
\sum_{\beta\ne\gamma}^3\Psi_{\beta}(\vec{x}_{\beta},\vec{y}_{\beta}) \;\;, 
\label{eq:eq} 
\end{equation}  
where $V^c_{\beta}$ is the Coulomb force between the particles of the pair 
denoted by channel $\beta$ ($\beta$=1,2,3) and $V^s_\gamma$ is the 
short-range pair interaction in the channel $\gamma$ ($\gamma$=1,2,3). 
The kinetic energy operator is given by 
$H_0=-\Delta_{\vec{x}_{\gamma}}-\Delta_{\vec{y}_{\gamma}}$ and $E$ is the 
total energy. The variables ${\vec{x}_{\gamma}},{\vec{y}_{\gamma}}$ 
are properly normalized relative Jacobi coordinate vectors, as defined 
in Ref.\cite{FGS03}. The wave function of the three-body system 
$\Psi$ is given by a sum over the three Faddeev components, 
$\Psi =\sum^3_{\gamma=1}\Psi_\gamma$. 

When two particles of the three-body system are identical, as the 
$\alpha$ particles are in $^9_{\Lambda}$Be ($\alpha\alpha\Lambda$), 
the coupled set of Faddeev equations reduces to two equations: 
\begin{equation} 
\begin{array}{l} 
(H_0+V_{\alpha\alpha}+V^c_3-E)U_3=-V_{\alpha\alpha}(U_1+P_{12}U_1) \;\;, \\ 
(H_0+V_{\alpha\Lambda}+V^c_1-E)U_1=-V_{\alpha\Lambda}(U_3+P_{12}U_1) \;\;, 
\end{array} 
\label{eq:fad} 
\end{equation} 
where $P_{12}$ is the permutation operator for the 
$\alpha$ bosons (particles 1,2), $V_{\alpha\alpha}$ and 
$V_{\alpha\Lambda}$ are nuclear potentials for the $\alpha\alpha$ and 
$\alpha\Lambda$ interactions respectively, $U_3$ is the Faddeev 
component corresponding to the rearrangement channel 
$(\alpha\alpha)-\Lambda$ and $U_1$ corresponds to the 
rearrangement channel $(\alpha\Lambda)-\alpha$, $V^c_\gamma$ is 
the Coulomb potential for channels $\gamma=1,3$. The total 
wave function is expressed by the components $U_1$ and $U_3$: $ 
\Psi = U_3 + (1 + P_{12})U_1$. The total orbital angular momentum 
is given by ${\vec L} = {\vec \ell}_{\alpha\alpha}+{\vec 
\lambda}_{(\alpha\alpha)-\Lambda} = {\vec 
\ell}_{\alpha\Lambda}+{\vec \lambda}_{(\alpha\Lambda)-\alpha}$, 
where ${\ell}_{\alpha\alpha}$ (${\ell}_{\alpha\Lambda}$) is the 
orbital angular momentum of the $\alpha\alpha$ 
($\alpha\Lambda$) pair and ${\lambda}_{(\alpha\alpha)-\Lambda}$ 
(${\lambda}_{(\alpha\Lambda)-\alpha}$) is the orbital angular 
momentum of the $\Lambda$ hyperon ($\alpha$ particle) with respect 
to the center of mass of the $\alpha\alpha$ ($\alpha\Lambda$) pair. 
Bose symmetry for the $\alpha\alpha$ pair requires that 
${\ell}_{\alpha\alpha}$ assumes non-negative even values, and the 
positive parity of the $_{\Lambda}^9$Be states considered here 
requires that ${\ell}_{\alpha\alpha}+{\lambda}_{(\alpha\alpha)-\Lambda}$ 
and ${\ell}_{\alpha\Lambda}+{\lambda}_{(\alpha\Lambda)-\alpha}$ are 
non-negative even numbers. For convenience we have assembled in the 
Appendix of the present paper details related to the partial-wave 
decomposition of Eqs.(\ref{eq:fad}). 

The lowest states of $_{\Lambda}^9$Be are expected to follow 
the structure of the corresponding low-lying states of $^8$Be. 
Thus, the $(\frac12)^+$ ground state of $_{\Lambda}^9$Be, 
considered as an $\alpha\alpha\Lambda$ system, is assumed 
to have $L=0$ total orbital angular momentum. 
The allowed combinations of the relative angular momenta 
(${\ell}_{\alpha\alpha}$, ${\lambda}_{(\alpha\alpha)-\Lambda}$) 
are (0,0), (2,2), (4,4), $\dots$; and for (${\ell}_{\alpha\Lambda}$, 
${\lambda}_{(\alpha\Lambda)-\alpha}$) are (0,0), (1,1), (2,2), (3,3), 
(4,4), $\dots$. The quantum numbers ${\ell}_{\alpha\alpha}$ 
and ${\ell}_{\alpha\Lambda}$ specify completely the orbital 
angular momentum states of the corresponding subsystems. 

The first excited $({\frac32}^+,{\frac52}^+)$ doublet of 
$_{\Lambda}^9$Be is assumed to have $L=2$ total orbital 
angular momentum. The allowed combinations of the relative 
angular momenta (${\ell}_{\alpha\alpha}$, 
${\lambda}_{(\alpha\alpha)-\Lambda}$) are (0,2), (2,0), (2,2), (2,4), 
(4,2), (4,4), $\dots$; and for (${\ell}_{\alpha\Lambda}$, 
${\lambda}_{(\alpha\Lambda)-\alpha}$) are (0,2), (1,1), (1,3), (2,0), 
(2,2), (2,4), $\dots$.

For the $\Lambda\Lambda\alpha$ system the Faddeev equations (\ref{eq:fad}) 
have a similar form (but with a minus sign in front of $P_{12}$ for the 
$\Lambda$ fermions, and without the Coulomb potential) which has been 
given in Refs.\cite{FGa02b,FGS03} for the $^{~~6}_{\Lambda\Lambda}$He 
hypernucleus.

\section{Potentials}
\label{sec:pot} 

\begin{figure}
\epsfig{file=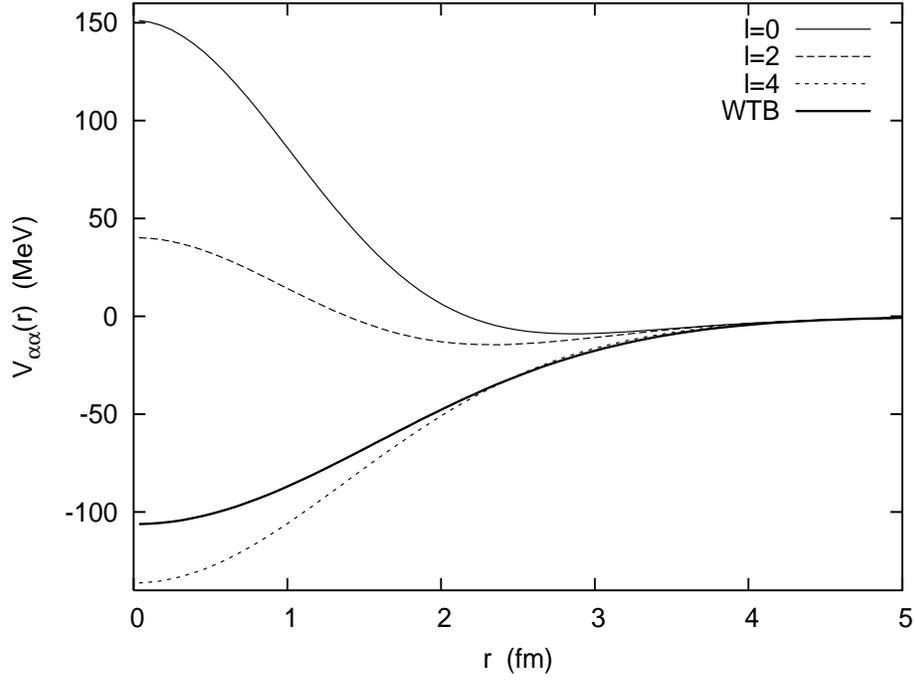, height=90mm, width=120mm}
\vspace*{3mm} 
\caption{$\alpha\alpha$ potentials.} 
\label{fig:fgs041} 
\end{figure}

To describe interactions in the $\alpha \Lambda \Lambda$ and 
$\alpha \alpha \Lambda$ systems, local pairwise potentials are used. 
The input $\alpha \alpha$ nuclear interaction is 
given by version $a$ of the phenomenological Ali-Bodmer (AB) potential 
\cite{AB66} (as modified in Ref.\cite{FJ96}) or by the Chien-Brown 
(CB) potential \cite{CB74} which include $s$, $d$ and $g$-wave components, 
fitted to the low-energy phase shifts in the $\alpha \alpha$ system. 
The resulting $\alpha \alpha$ potential has the following form:
\begin{equation}
\label{eq:aa}
V_{\alpha\alpha}(r) =  \sum_{l=0,2,4} V^l_{\alpha\alpha}(r)P_l \;\;, 
\end{equation}
where $P_l$ is a projector onto the state of the $\alpha \alpha$
pair with orbital angular momentum $l$. The $s$-wave component
$V^0_{\alpha\alpha}(r)$ has a strong repulsive core which
simulates a Pauli blocking effect for the $\alpha \alpha$ pair 
at short distances. The $l$ dependence of the CB potential is 
demonstrated in Fig. \ref{fig:fgs041}, where we added for comparison 
an $l$-independent potential (WTB) due to Ref.\cite{WTB86}. This latter 
potential is purely attractive, yet at short distances the attraction 
is moderated by the Coulomb repulsion between the two $\alpha$s which 
is always included in the present calculation. 

For the $\alpha\Lambda$ interaction, several potentials used in previous 
calculations \cite{MBI83,PC91,OKS00,CFK02} are listed in Table 
\ref{tab:tabl1}. These potentials have different shapes, while reproducing 
closely the experimental value of the binding energy for the $^5_\Lambda$He 
hypernucleus \cite{Dav91} which is considered to be an $s$-wave bound state 
of the $\alpha\Lambda$ system. The Tang-Herndon (TH) potential \cite{The65} 
used in Ref.\cite{OKS00} and the Gibson I (Gibson) potential \cite{GGW72} 
used in Ref.\cite{DGN82} are purely attractive Gaussian $s$-wave potentials. 
The Maeda-Schmid $s$-wave potential (MS) \cite{MS84} is a sum of two 
Woods-Saxon functions, where the inner function is repulsive, moderating at 
short distances the attraction due to the outer function:
\begin{equation}
\label{eq:WS}
V_{\alpha\Lambda}(r)=\frac{V_{\rm rep}}{1+\exp((r-R_{\rm rep})/a_{\rm rep})} 
-\frac{V_{\rm att}}{1+\exp((r-R_{\rm att})/a_{\rm att})} \;\;. 
\end{equation} 
The Isle $s$-wave potential \cite{KAT85} is a sum of two Gaussians 
where the inner Gaussian is repulsive, outweighing at short distances 
the attractive outer Gaussian. In addition to these $s$-wave 
$\alpha\Lambda$ potentials, we used for the first time ever in 
$^9_{\Lambda}$Be binding energy calculations a potential (MSA) proposed 
in Ref.\cite{MSA03} that includes also a $p$-wave component inspired by 
the NSC97 model of the $\Lambda N$ interaction. 
Generally, the contribution of the various partial 
waves to the $\alpha\Lambda$ potential is given by: 
\begin{equation} 
\label{eq:aL} 
V_{\alpha\Lambda}(r) = \sum_{l=0,1,\dots} V^l_{\alpha\Lambda}(r)P_l \;\;, 
\end{equation} 
where $P_l$ is a projector onto the state of the $\alpha \Lambda$ 
system with orbital angular momentum $l$, and 
\begin{equation}
\label{eq:Isle} 
V^l_{\alpha\Lambda}(r) =
V^l_{\rm rep}\exp(-(r/\beta^l_{\rm rep})^2) - 
V^l_{\rm att}\exp(-(r/\beta^l_{\rm att})^2) \;\;, 
\end{equation} 
are the partial-wave components of the potential. The calculated 
$s$-wave scattering lengths and effective ranges, and the 
$^5_{\Lambda}$He binding energy for all of these potentials, 
are listed in Table \ref{tab:tabl2}. 

\begin{figure}
\epsfig{file=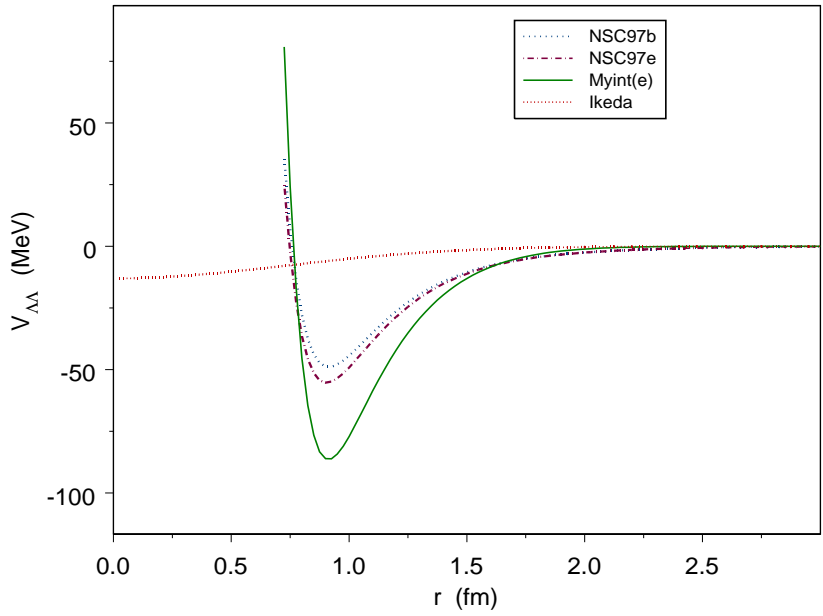, height=90mm, width=120mm}
\vspace*{3mm}
\caption{$\Lambda\Lambda$ potentials.}
\label{fig:fgs042}
\end{figure}

The $\Lambda\Lambda$ interaction potentials in the $^{1}S_0$
channel which are used as input to the Faddeev equations are
of a three-range Gaussian form
\begin{equation}
\label{eq:HKM}
V_{\Lambda\Lambda} = \sum_i^3 v^{(i)}\exp(-r^2/\beta_i^2)\;\;,
\end{equation}
following the work of Hiyama et al. \cite{HKM97} where a
phase-equivalent $\Lambda\Lambda$ potential of this soft-core form
was fitted to the Nijmegen model D (ND) hard-core interaction
\cite {ND75} assuming the same hard core for the $NN$ and
$\Lambda\Lambda$ potentials in the $^{1}S_0$ channel. For other
interactions, notably the Nijmegen soft-core NSC97 model
$\Lambda\Lambda$ potentials \cite{SRi99}, we have renormalized the
strength of the medium-range attractive component ($i=2$) of this
potential fitting as closely as possible the scattering length and
the effective range. The appropriate range and strength parameters
are listed in Tables 1 and 2 of Ref. \cite{FGa02b}. For the NSC97e
interaction, Myint et al. \cite{MSA03} have used a different
parameterization which is listed in Table 3 of their paper. 
These soft-core-repulsion $\Lambda\Lambda$ potentials, 
as well as a purely attractive Gaussian potential 
$V_{\Lambda\Lambda}={\frac14}(-52.25\exp(-r^2/1.034^2))$, 
are shown in Fig. \ref{fig:fgs042}. This latter potential (marked Ikeda) 
was obtained by taking 1/4 of the potential used by Ikeda et al. 
\cite{IBM85}.  

\section{Results} 
\label{sec:res} 

The Faddeev coupled equations (\ref{eq:fad}) for the $\alpha\Lambda\Lambda$ 
and $\alpha\alpha\Lambda$ systems were solved numerically in configuration 
space, using the method given in Ref. \cite{BSS96} and applied by us in 
Ref. \cite{FGS03}. The partial-wave decomposition of these equations 
is relegated to the Appendix. We applied different values for the cutoff 
radius used for the $\alpha\Lambda\Lambda$ and $\alpha\alpha\Lambda$ systems.
For the $\alpha\Lambda\Lambda$ system a cutoff radius value 
$\rho_{\rm cutoff}=25$ fm was used, whereas for the $\alpha\alpha\Lambda$ 
system a value $\rho_{\rm cutoff}=40$ fm was used (considering the long-range 
Coulomb force). 

\subsection{$^{~~6}_{\Lambda\Lambda}$He} 
\label{subsec:He}

\begin{figure}
\epsfig{file=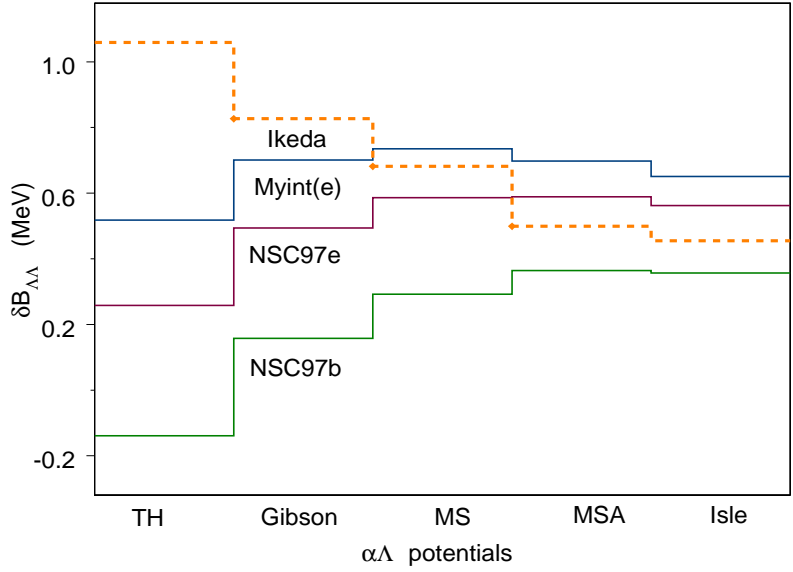, height=90mm, width=120mm}
\vspace*{3mm}
\caption{The $\Lambda\Lambda$ pairing energy 
$\delta B_{\Lambda\Lambda}$, Eq. (\ref{eq:deltal}), 
for several $\Lambda\Lambda$ and $\alpha\Lambda$ potentials.} 
\label{fig:fgs043} 
\end{figure}

The $\Lambda\Lambda$ binding energies $E_B$ 
($E_{B} = - B_{\Lambda\Lambda}$) calculated for the $0^+$ 
ground state of $^{~~6}_{\Lambda\Lambda}$He, viewed as a three-body system 
$\alpha\Lambda\Lambda$ with $L^{\pi}=0^+, S=0$ quantum numbers, 
are listed in Table \ref{tab:tabl3} for several combinations of 
$\Lambda\Lambda$ and $\alpha\Lambda$ interaction potentials 
discussed in Sect. \ref{sec:pot}. The orbital angular momenta included 
are $l_{\alpha\Lambda}$=0,1,2,3,4,5,6, $l_{\Lambda\Lambda}$=0,2,4,6, 
ensuring convergence within few keV \cite{FGS03}. 
Here an `$s$-wave' model is used for the $\alpha\Lambda$ interaction, 
meaning that $V^{l}_{\alpha\Lambda}$ is independent of $l$ and equals 
$V^{l=0}_{\alpha\Lambda}$. The high partial-wave contributions of the 
$\alpha\Lambda$ potential for this system are small and in total do not 
exceed about 0.2 MeV \cite{FGS03}. 
Note that for $V_{\Lambda\Lambda}=0$, the calculated binding energy is 
larger (in absolute values) than twice the $^5_{\Lambda}$He binding energy 
for all of the $\alpha\Lambda$ potentials listed in Table \ref{tab:tabl2}. 
Hence $\Delta B_{\Lambda\Lambda}(V_{\Lambda\Lambda}=0)>0$. 
This nonzero value is due to the mass-polarization term contained in the 
kinetic energy operator \cite{FGa02a,HKM02}. In order to eliminate the 
contribution from this term, we have listed in Table \ref{tab:tabl3}, 
in parentheses, the $\Lambda\Lambda$ `pairing' energies corresponding 
to a given $V_{\Lambda\Lambda}$, using the definition 
\begin{equation} 
\label{eq:deltal} 
\delta B_{\Lambda\Lambda}=E_B(V_{\Lambda\Lambda}=0)-E_B(V_{\Lambda\Lambda})\;. 
\end{equation} 
The calculated $\delta B_{\Lambda\Lambda}$ values are also plotted in 
Fig. \ref{fig:fgs043}. For a given $\Lambda\Lambda$ interaction, 
a clear dependence of $\delta B_{\Lambda\Lambda}$ on the $\alpha\Lambda$ 
potential is observed, and it is related to the hard core of the 
$\Lambda\Lambda$ interaction which acts more effectively for purely 
attractive $\alpha\Lambda$ potentials (TH, Gibson) that `compress' 
the three-body system, leading to a smaller value of 
$\delta B_{\Lambda\Lambda}$ than for $\alpha\Lambda$ repulsive-core 
potentials (MS, MSA, Isle). This dependence on the $\alpha\Lambda$ 
potential is reversed, as shown by the added dashed-line 
histogram in Fig. \ref{fig:fgs043}, for the purely attractive (Ikeda) 
$\Lambda\Lambda$ interaction potential. Anticipating some kind of 
repulsive cores in the $\Lambda\Lambda$ and $\alpha\Lambda$ 
interaction potentials, one may conclude that the uncertainty in 
$\delta B_{\Lambda\Lambda}$ due to the type of the $\alpha\Lambda$ 
potential is less than 0.1 MeV. We note that the uncertainty in 
$\Delta B_{\Lambda\Lambda}$ is larger, amounting for the Myint (e) 
potential in Table \ref{tab:tabl3} to 0.4 MeV which is close to the 
uncertainty found by Carr et al. \cite{CAG97} in their 
$\alpha\Lambda\Lambda$ study of $^{~~6}_{\Lambda\Lambda}$He 
(surveying a range of considerably stronger $\Lambda\Lambda$ 
interactions prior to the NAGARA event \cite{Tak01} report). 

\subsection{$^9_{\Lambda}$Be} 
\label{subsec:Be} 

The binding energies $E_B$ calculated for the $({\frac{1}{2}})^+$ 
ground state of $^9_{\Lambda}$Be, viewed as a three-body system 
$\alpha\alpha\Lambda$, are listed in Table \ref{tab:tabl4} for 
several combinations of $\alpha\alpha$ and $\alpha\Lambda$ 
interaction potentials discussed in Sect. \ref{sec:pot}. 
We have studied two different potential models. In the first one, 
the `$s$-wave' model as described above for $^{~~6}_{\Lambda\Lambda}$He, 
the $s$-wave $\alpha\Lambda$ potential acts in all partial waves of 
the $\alpha\Lambda$ subsystem: 
$V^l_{\alpha\Lambda}(r)=V^0_{\alpha\Lambda}(r)$, with $l$=1,2,3,4,5. 
In the second model, the `$s$ and $p$-wave' model, we retained only 
the $l=0,1$ partial-wave components (different from each other) of 
the MSA $\alpha\Lambda$ potential (\ref{eq:aL}). For the first model, 
the agreement between our results (FGS) and those of Cravo et al. 
\cite{CFK02} which were derived solving integral equations is 
sufficiently good. Nevertheless it should be noted that in the 
latter approach the nuclear potentials are approximated by separable 
potentials which might explain some small differences 
between our results and those of Ref. \cite{CFK02}. 

All partial waves up to $l=$5 were taken into 
account in our calculations, with the partial waves 
$l_{\alpha\Lambda}\leq 2$, $l_{\alpha\alpha}\leq 2$ dominating the 
total contribution. The Coulomb repulsion between the $\alpha$ 
clusters is included. The results are insensitive to which realistic 
$\alpha\alpha$ interaction (AB or CB) is used. Comparing the first 
and the second models with each other one observes that in the 
first model, the `$s$-wave' model, the $p$-wave contribution to the 
$\alpha\alpha\Lambda$ binding energy is substantial, exceeding 
1 MeV for the repulsive-core MSA and Isle potentials. Using 
the more realistic MSA $\alpha\Lambda$ potential \cite{MSA03} 
in the second model, the `$s$ and $p$-wave' model, with a weaker 
$p$-wave component, the calculated binding energy of $^9_{\Lambda}$Be 
is reduced significantly by about 0.75 MeV.  
The fairly wide spectrum of calculated binding energies for the 
$\alpha\alpha\Lambda$ system in the `$s$-wave' model, 
from 6 MeV to 8 MeV, may be explained in a similar way to the 
explanation of the considerably weaker dependence on the 
$\alpha\Lambda$ potential of the calculated binding energies 
for the $\alpha\Lambda\Lambda$ system in Table \ref{tab:tabl3}. 
The purely attractive TH and Gibson potentials `compress' the system, 
making the strongly repulsive core of the $s$-wave $\alpha\alpha$ 
interaction quite effective. Therefore, the calculated binding energy 
for these potentials is smaller (in absolute value) than for the 
Isle and MSA potentials. On the other hand, using a purely attractive 
unrealistic $\alpha\alpha$ potential (WTB, first row of 
Table \ref{tab:tabl4}), 
the dependence on the $\alpha\Lambda$ potential is negligible, 
not exceeding 0.1 MeV, except for MSA which deviates by 0.3 MeV. 

\begin{figure} 
\epsfig{file=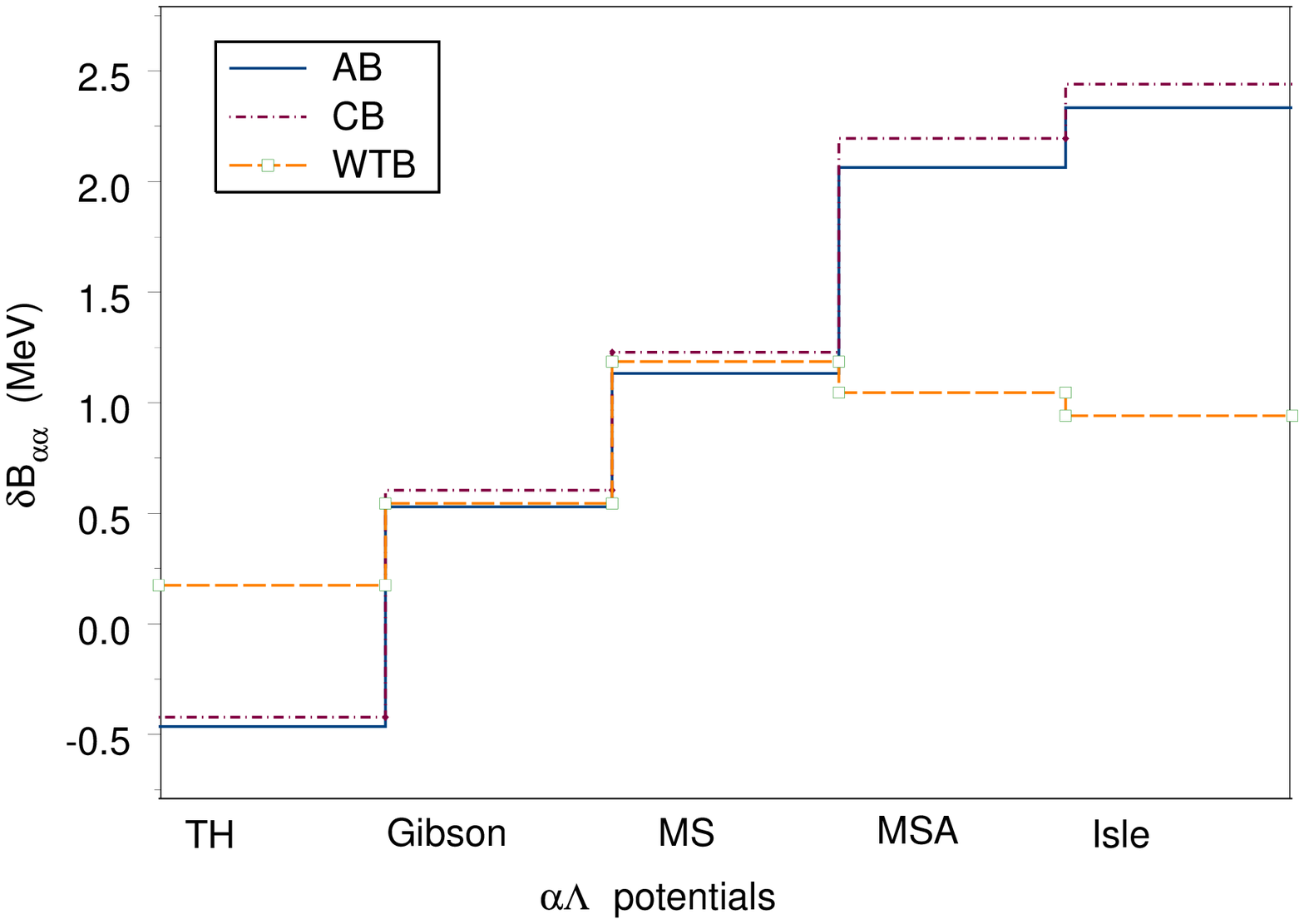, height=90mm, width=120mm} 
\vspace*{3mm} 
\caption{The $\alpha\alpha$ pairing energy 
$\delta B_{\alpha\alpha}$, Eq. (\ref{eq:deltaa}), 
for the AB and CB $\alpha\alpha$ potentials, for several 
$\alpha\Lambda$ potentials. The orbital angular 
momenta included in these calculations are 
$l_{\alpha\Lambda}$=0,1,2, and $l_{\alpha\alpha}$=0,2,4.}
\label{fig:fgs044} 
\end{figure}

The effect of the $\alpha\alpha$ interaction may be demonstrated 
by switching it off, while keeping on the Coulomb repulsion in 
these $\alpha\alpha\Lambda$ binding energy calculations. 
The corresponding binding energies given in the last row of 
Table \ref{tab:tabl4} vary by less than 1 MeV of each other, 
but in reverse order to that when $V_{\alpha\alpha}$ is on. 
Consequently, if one defines in analogy to Eq. (\ref{eq:deltal}) 
the $\alpha\alpha$ `pairing' energy corresponding to a given 
$V_{\alpha\alpha}$, 
\begin{equation} 
\label{eq:deltaa} 
\delta B_{\alpha\alpha}=E_B(V_{\alpha\alpha}=0)-E_B(V_{\alpha\alpha})\;, 
\end{equation} 
it displays quite a strong dependence on the $\alpha\Lambda$ interaction 
used in the calculation, extending over a range of 3 MeV. 
This is shown in Fig.\ref{fig:fgs044}. 
The attractive nature of the $\alpha\alpha$ interaction comes out 
clearly for $\alpha\Lambda$ interaction potentials that contain inner 
repulsion. The magnitude of $\delta B_{\alpha\alpha}$ is modest with 
respect to what it would have been, had the Pauli principle been 
implicitly ignored in the construction of the $\alpha\alpha$ potentials. 
There is hardly a difference between 
the $\alpha\alpha$ potentials AB and CB, whereas the variation of 
$\delta B_{\alpha\alpha}$ for the purely attractive WTB $\alpha\alpha$ 
potential is more moderate. In summary, $^9_{\Lambda}$Be, 
with a relatively light $\Lambda$ `core' particle, does not provide 
a useful hadronic medium in which to determine the $\alpha\alpha$ 
pairing energy. 

Finally, in Table \ref{tab:tabl5} we show the calculated excitation 
energies of the excited (${\frac 32}^+,{\frac 52}^+$) spin-flip 
doublet of $^9_{\Lambda}$Be levels based on the first excited 
$2^+$ $^8$Be level. The agreement between our results and those by 
Cravo et al. \cite{CFK02}, and also with respect to the observed 
levels which are split by less than 50 keV \cite{AAC02}, is particularly 
good for the choice of CB $\alpha\alpha$ interactions. In contrast, 
the non-Faddeev calculations by Portilho and Coon \cite{PC91} 
overestimate substantially the Faddeev results and should be looked 
at with a grain of salt. 
In our calculation, $l_{\alpha\Lambda}=0,1,2$ partial waves are 
included, except for the `$s$ and $p$-wave' results which used 
$l_{\alpha\Lambda}=0,1$, together with $l_{\alpha\alpha}=0,2,4$ 
everywhere.

\section{Summary} 
\label{sec:sum}

The main question addressed in this work was the model dependence 
of $\delta B_{\Lambda\Lambda}$, extracted from the calculated 
$^{~~6}_{\Lambda\Lambda}$He binding energy, due to the 
$\alpha\Lambda$ potential used in the configuration-space 
Faddeev $\alpha\alpha\Lambda$ calculation. We have found that this 
model dependence does not exceed 0.1 MeV, provided realistic 
$\alpha\Lambda$ potentials that consist of both short-range repulsion 
and longer-range attraction are used. The $\alpha\Lambda$ potential 
manifests itself in the Faddeev calculation mostly through its 
$s$-wave component which is regulated by fitting to $^5_{\Lambda}$He 
binding energy, the higher partial waves adding up to 0.2 MeV 
\cite{FGS03}. The dependence on the particular structure of the 
$p$-wave component is then less than 0.1 MeV. In contrast, the 
calculated $^9_{\Lambda}$Be binding energy is very sensitive to 
the $p$-wave contents of the $\alpha\Lambda$ potential. In agreement 
with Hiyama et al. \cite{HKM97} we find effects of up to 1 MeV 
due to the $p$-wave component of the $\alpha\Lambda$ interaction. 
We have found that the $p$-wave component must be weakened considerably 
with respect to what an $l$-independent $\alpha\Lambda$ potential 
constrained only by the $^5_{\Lambda}$He binding energy yields. 
Using the MSA $l$-dependent potential \cite{MSA03}, the calculated 
binding energy of $^9_{\Lambda}$Be is still overestimated by 0.6-0.7 
MeV, an amount which could possibly arise from three-body forces 
\cite{BUs87} unaccounted for in these Faddeev calculations.

\section{Appendix}

The partial-wave decomposition of Eq.(\ref{eq:eq}) for the 
$\alpha\alpha\Lambda$ system leads to the following set of partial
differential equations for the Faddeev components 
$\Psi_{\alpha}=\rho^{-1/2}U_{\alpha}$, in polar coordinates
$\rho^{2}=x_{\alpha}^{2}+y_{\alpha}^{2}$, $\tan\theta =\vert
y_{\alpha}\vert /\vert x_{\alpha}\vert$ :
\begin{equation}
\label{eq:aaL1}
\begin{array}{l}
\displaystyle  \{-\frac{\partial^{2}}{\partial \rho^{2}}
    -\frac{1}{\rho^{2}}\frac{\partial^{2}}{\partial \theta^{2}}
    +V^l_1(\rho,\theta) + \frac{l(l+1)}{\rho^{2}\cos^{2}\theta}+
\frac{\lambda (\lambda+1)}{\rho^{2}\sin^2\theta}
    -\frac{1}{4\rho^{2}}-E\}U^{l\lambda}_1(\rho,\theta)\\
\displaystyle 
+\frac{q}{\rho}(-1)^L\sum_{l^{\prime},\lambda^{\prime}}\Pi_{l\lambda
l^{\prime}\lambda^{\prime}}
U^{l^{\prime}\lambda^{\prime}}_1(\rho,\theta) 
 \sum_{\tau=\max(\vert l-l^{\prime}\vert,\vert \lambda-\lambda ^{\prime}\vert)}
^{\min(l+l^{\prime},\lambda+\lambda^{\prime})}\frac{b_{\tau}}{2\tau+1}
   C^{\tau0}_{l^{\prime}0l0}C^{\tau0}_{\lambda^{\prime}0\lambda0}
   \left \{ \begin{array}{ccc} 
\lambda     & \lambda ^{\prime} & \tau \\
   \l^{\prime} & l                 & L
   \end{array} \right \} \\ 
\displaystyle = -\frac{1}{2}V^l_1(\rho,\theta)
\sum_{l^{\prime},\lambda ^{\prime}}\{(-1)^{l^{\prime}}
(h^{L21}_{l\lambda, l^{\prime}\lambda^{\prime}}
U^{l^{\prime}\lambda^{\prime}}_{1})(\rho,\theta)+(-1)^{l+l^{\prime}}
(h^{L13}_{l\lambda, l^{\prime}\lambda^{\prime}}
U^{l^{\prime}\lambda^{\prime}}_{3})(\rho,\theta)\}, 
\end{array}
\end{equation} 

\begin{equation} 
\label{eq:aaL2} 
\begin{array}{l} 
\displaystyle
\{-\frac{\partial^{2}}{\partial \rho^{2}}
    -\frac{1}{\rho^{2}}\frac{\partial^{2}}{\partial \theta^{2}}
    +V^l_3(\rho,\theta) 
    + \frac{l(l+1)}{\rho^{2}\cos^{2}\theta}
    + \frac{\lambda (\lambda+1)}{\rho^{2}\sin^2\theta}
    +  \frac{q}{\rho\cos\theta}
    - \frac{1}{4\rho^{2}} - E\}U^{l\lambda}_3(\rho,\theta)\\
\displaystyle
    = -V^l_3(\rho,\theta) \sum_{l^{\prime},\lambda^{\prime}} 
      (h^{L31}_{l\lambda, l^{\prime}\lambda^{\prime}}
U^{l^{\prime}\lambda^{\prime}}_{1}) (\rho,\theta) \;, 
\end{array}
\end{equation}
where $\Pi_{l\lambda ...}=\sqrt{(2l+1)(2\lambda +1)\cdot\cdot\cdot}$, 
$C^{\tau0}_{l^{\prime}0l0}$ are Clebsch-Gordan coefficients 
and $\big\{\cdot \cdot \cdot\big\}$ is a $6j$ symbol. 
Here $b_{\tau}$ stand for 
$$
\begin{array}{l}
\displaystyle b_{\tau}=\left \{
\begin{array}{l}
\displaystyle
\frac{1}{\vert C_{31} \vert \cos\theta}\Big(\frac{S_{31}\sin\theta}
{C_{31}\cos\theta}\Big)^{\tau},
\ \ \ \ \tan\theta<\frac{C_{31}}{S_{31}},
\\
\displaystyle
\frac{1}{\vert S_{31} \vert \sin\theta}\Big(\frac{C_{31}\cos\theta}
{S_{31}\sin\theta}\Big)^{\tau},
\ \ \ \ \tan\theta\geq\frac{C_{31}}{S_{31}},
\end{array}
\right.
\end{array}
$$
and $q$ denotes the mass-scaled charge 
$$
q=4\frac{e^2}{\hbar}\sqrt{m_\alpha m_\Lambda}.
$$
We used the scaled Jacobi vectors ${\vec{x}_{\alpha},\vec{y}_{\alpha}}$, 
related to the standard particle vector coordinates by   
\begin{equation}
\label{eq:xy}
\begin{array}{l}
\vec{x}_{\alpha}=\left(\frac{2m_{\beta}m_{\gamma}}{m_{\beta}+
m_{\gamma}}\right)^{1/2}(\vec{r}_{\beta}-\vec{r}_{\gamma}), \\
\vec{y}_{\alpha}=\left(\frac{2m_{\alpha}(m_{\beta}+
m_{\gamma})}{M}\right)^{1/2}(\frac{m_{\beta}\vec{r}_{\beta}+
m_{\gamma}\vec{r}_{\gamma}}{m_{\beta}+m_{\gamma}}-\vec{r}_{\alpha}),
\end{array} 
\end{equation} 
as independent coordinates. Here ($\alpha,\beta,\gamma$) is a cyclic 
permutation of (1,2,3) and $M$ is the total mass. The Jacobi vectors for 
different $\alpha$'s are linearly related by an orthogonal transformation: 
$$
  \left(
  \begin{array}{c}
     \vec{x}_{\alpha} \\ \vec{y}_{\alpha}
  \end{array}
  \right)=
  \left(
  \begin{array}{rl}
      C_{\alpha\beta} & S_{\alpha\beta} \\
     -S_{\alpha\beta} & C_{\alpha\beta}
  \end{array}
  \right)
  \left(
  \begin{array}{c}
     \vec{x}_{\beta} \\ \vec{y}_{\beta}
  \end{array}
  \right) \ ,\ \ \ C^2_{\alpha\beta} + S^2_{\alpha\beta} = 1,
$$
where
$$
C_{\alpha\beta}=-\sqrt{\frac{m_{\alpha}m_{\beta}}
{(M-m_{\alpha})(M-m_{\beta})}} \ \ , \ \ S_{\alpha\beta} =
(-)^{\beta - \alpha}sgn(\beta - \alpha)
\sqrt{1-C^{2}_{\alpha\beta}}.
$$
In a bispherical basis, the kernel 
$h^{L\beta\alpha}_{l\lambda,l^{\prime}\lambda^{\prime}}$ 
of the integral operator in Eqs.(\ref{eq:aaL1},\ref{eq:aaL2}) has the form
\begin{equation}
\label{eq:A1}
\displaystyle (h^{L\beta\alpha}_{l\lambda,
l^{\prime}\lambda^{\prime}}
  U_{\beta(\alpha)}^{l^{\prime}\lambda^{\prime}})(\rho,\theta) 
  = \int_{-1}^{+1}du
  \frac{\sin\theta \cos\theta}{\sin\theta^{'} \cos\theta^{'}}
  h^{L\beta\alpha}_{l\lambda, l^{\prime}\lambda^{\prime}}(\theta,
  \theta^{\prime}(u,\theta))
U_{\beta(\alpha)}^{l^{\prime}\lambda^{\prime}}(\rho,\theta^{\prime}(u,\theta)),
\end{equation}
where
$$
\cos^{2}\theta^{'}(u,{\theta}) =
C^{2}_{\beta\alpha}\cos^{2}\theta
 +2C_{\beta\alpha}S_{\beta\alpha}\cos\theta \sin\theta\cdot u 
+S^{2}_{\beta\alpha}\sin^{2}\theta.
$$ 
More explicitly: 
\begin{equation}
\label{eq:A2} 
\begin{array}{l}
\displaystyle 
   h^{L\beta\alpha}_{l\lambda,l^{\prime}\lambda^{'}}(\theta,\theta^{'}) 
   = (-)^{L+l^{\prime}+\lambda^{\prime}}\Pi_{l \lambda}
   \Pi^2_{l^{\prime}\lambda^{\prime}}
   \sqrt{(2l^{\prime})!(2\lambda^{\prime})!} \\ 
\displaystyle \times \sum_{
   \lambda_{1}+\lambda_{2}=\lambda^{'} \;\; l_{1}+l_{2}=l^{'}} 
   \frac{\sin^{\lambda_{1}+l_{1}}\theta \cos^{\lambda_{2}+l_{2}}\theta}
   {\sin^{\lambda^{'}}\theta^{'} \cos^{l^{'}}\theta^{'}}
   \frac{(-)^{\lambda_2}C^{\lambda_{1}+l_{2}}_{\beta\alpha}
   S^{\lambda_{2}+l_{1}}_{\beta\alpha}} 
   {((2\lambda_{1})!(2\lambda_{2})!(2l_{1})!(2l_{2})!)^{1/2}}\\ 
\displaystyle \times \; \sum_{ \lambda^{"}\;l^{"}}
   \Pi^2_{l^{''}\lambda^{''}}
   \left ( \begin{array}{ccc} 
   \lambda_{1} & l_{1} & \lambda^{''} \\
    0          & 0     & 0
   \end{array} \right)
   \left ( \begin{array}{ccc} 
  \lambda_{2} & l_{2} & l^{''} \\
    0          & 0     & 0
   \end{array} \right) 
   \left \{ \begin{array}{ccc}
   \lambda_{1} & \lambda_{2} & \lambda^{'} \\
    l_{1}      &  l_{2}      &  l^{'} \\
   \lambda^{''} &  l^{''}      &  L
   \end{array} \right \} \\      
\displaystyle \times \;
\sum_{k=0}(-)^{k}(2k+1) 
   \left ( \begin{array}{ccc}
    k          & \lambda^{''} & \lambda \\
    0          & 0           & 0
   \end{array} \right) 
   \left ( \begin{array}{ccc}
    k          & l^{''} & l \\
    0          & 0     & 0
   \end{array} \right)  
\left \{ \begin{array}{ccc}
    l          & \lambda & L \\
   \lambda^{''} & l^{''}   & k
   \end{array} \right \} 
P_{k}(u)\;, 
\end{array}
\end{equation}
in terms of $3j$, $6j$, $9j$ symbols and Legendre polynomials $P_k(u)$. 
The index $k$ runs in Eq. (\ref{eq:A2}) from zero to
$(\lambda^{'}+l^{'}+\lambda+l)/2$. 

For zero total orbital angular momentum $L=0$ 
($\lambda= l,\lambda^{'}= l^{'}$), all the summations in the 
expression above may be carried out to obtain a simpler expression 
of the form 
$$
h^{\beta\alpha}_{ l l^{'}}(\theta,\theta^{'})=(-1)^{l+l^{'}}\sqrt{
(2l+1)(2l^{'}+1)}P_l(u)P_{l^{'}}(u^{'})\;,
$$
where 
$$
u^{'}=
\frac{-\cos(2\theta)+(C^{2}_{\alpha\beta}-S^{2}_{\alpha\beta})\cos(2\theta^{'})}
{2C_{\alpha\beta}S_{\alpha\beta}\sin(2\theta^{'})}\;.
$$

To solve the eigenvalue problem in the region $\rho\in[0,\infty], 
\theta \in[0,\pi/2]$, Eqs.(\ref{eq:aaL1},\ref{eq:aaL2}) must be 
supplemented by the boundary conditions
$$
\begin{array}{l}
    U^l_{\gamma}(0,\theta) = U^l_{\gamma}(\infty,\theta) =
U^l_{\gamma}(\rho,0) = U^l_{\gamma}(\rho,\pi/2) = 0.
\end{array}
$$
In Eqs.(\ref{eq:aaL1},\ref{eq:aaL2}) the Coulomb repulsion between the $\alpha's$ 
is included rigorously, allowing full account of any number of 
partial waves. In actual calculations, the number of terms in the 
summation over $l^{\prime}$ is truncated once convergence has been achieved.

\begin{acknowledgments}
This work was partially supported by the Israel Science Foundation
(grant 131/01). The work of I.F and V.M.S was supported by the
RFFI under Grant No. 02-02-16562.
\end{acknowledgments}

\newpage

\begin{table}
\caption{Parameters of $\alpha\Lambda$ potentials used in the present 
work. The notations follow Eq.(\ref{eq:Isle}), except for MS where 
$\beta=(R,a)$ in terms of Eq.(\ref{eq:WS}).} 
\label{tab:tabl1} 
\begin{center} 
%\begin{ruledtabular} 
\begin{tabular}{lccccc}
 Potential & $l$ & $V_{\rm rep}^l$ (MeV) & $\beta_{\rm
rep}^l$ (fm) & $V_{\rm att}^l$ (MeV) & $\beta_{\rm att}^l$ (fm)\\
\hline TH \cite{OKS00}       & 0 & --    & --   & 60.17 & 1.2729 \\
       Gibson \cite{DGN82}  & 0 & --    & --   & 43.48 & 1.5764 \\
       MS \cite{MS84} & 0 & 18.09 & (0.88, 0.2353) & 35.98 & (1.72, 0.3541) \\
       Isle(DA) \cite{KAT85} & 0 & 450.4 & 1.25 & 404.9 & 1.41 \\
       MSA \cite{MSA03}      & 0 & 91.0  & 1.3  & 95.0  & 1.7 \\
                             & 1 & 33.4  & 1.3  & 39.4  & 1.7 \\
\end{tabular}
%\end{ruledtabular}
\end{center} 
\end{table}

\begin{table}
\caption{Scattering length $a$, effective range $r_0$ and binding 
energy $E_B$ of the $_\Lambda^5$He hypernucleus for various 
$\alpha\Lambda$ potentials.} 
\label{tab:tabl2}
\begin{center} 
%\begin{ruledtabular}
\begin{tabular}{lccc}
Potential & $a$ (fm) & $r_0$ (fm) &$ E_B$\\ \hline
Isle(DA) & 4.24  & 2.05& -3.10\\
MSA      & 4.18  & 1.97& -3.12\\
MS       & 4.00  & 1.67& -2.84\\
Gibson   & 3.80  & 1.53&-3.08\\
TH       & 3.63  & 1.32& -3.03\\
\hline exp. \cite{Dav91} &  --  & --   & -3.12$\pm$0.02
\end{tabular}
%\end{ruledtabular}
\end{center} 
\end{table} 

\begin{table}
\caption{$E_{B}(_{\Lambda\Lambda}^{~~6}$He) (and in parentheses 
$\delta B_{\Lambda\Lambda}$, Eq. (\ref{eq:deltal})) in MeV, calculated  
for various $\Lambda\Lambda$ and $\alpha\Lambda$ potentials. Energy 
is measured with respect to the $\alpha + \Lambda + \Lambda$ threshold. 
The `s-wave' model is used for the $\alpha\Lambda$ interaction. 
$E_B^{\rm exp}(0^+)=-7.25 {\pm 0.19}^{+0.18}_{-0.11}~{\rm MeV}$ 
{\protect \cite{Tak01}}.} 
\label{tab:tabl3}
\begin{center}
%\begin{ruledtabular}
\begin{tabular}{cccccc}
 Potential     &  TH      & Gibson  &  MS     & MSA     & Isle  \\
\hline
Myint(e)       & -6.853   & -7.084  & -6.853  & -7.107  & -6.992 \\
               & (0.518)  &  (0.701)& (0.735) &  (0.698)&   (0.651)   \\
\hline
NSC97e         & -6.593   & -6.877  & -6.476  & -6.998  & -6.903 \\
               & (0.258)  & (0.494) & (0.586) & (0.589) &  (0.562)
\\  \hline
NSC97b         & -6.200   &-6.541   & -6.182  & -6.773  & -6.698  \\
               & (-0.139) & (0.158) & (0.292) & (0.364) & (0.357) \\  
\hline
$V_{\Lambda\Lambda}$=0& -6.335 & -6.383  & -5.890 & -6.409 & -6.341
\end{tabular}
%\end{ruledtabular}
\end{center}
\end{table}

\begin{table}
\caption{$E_B(_{\Lambda}^{9}$Be g.s.) in MeV, calculated 
for various ${\alpha\alpha}$ and $\alpha\Lambda$ potentials. 
Energy is measured with respect to the ${\alpha+\alpha+\Lambda}$
threshold. $E_B^{\rm exp}({\frac{1}{2}}^+)=-6.62 \pm 0.04$ MeV 
{\protect \cite{Dav91}}.} 
\label{tab:tabl4}
\begin{center} 
%\begin{ruledtabular}
\begin{tabular}{cccccccccc}
Ref. &${\alpha\alpha}$ & ${\alpha\Lambda}$ & $l_{\alpha\Lambda}$ &
$l_{\alpha\alpha}$ &
                                       TH  & Gibson&  MS  &  MSA  &
Isle(DA) \\
\hline
FGS &WTB& `$s$-wave' &0,1,2      & 0,2,4  &-6.623 &-6.725 &-6.718 & 
-6.932 & -6.726\\
\hline 
FGS &AB& `$s$-wave' & 0          & 0      &-5.043 &-5.598 &-5.530 & 
-6.542 & -6.581\\
&  &            & 0          & 0,2,4  &-5.553 &-5.976 &-5.836 &-6.749 &
-6.804\\
&  &            & 0,1        & 0,2,4  &-5.924 &-6.606 &-6.571 &-7.851 &
-8.030\\
&  &            & 0,1,2      & 0,2,4  &-5.991 &-6.709 &-6.664 &-7.947 &
-8.119\\
&  &            & 0,1,2,3,4,5& 0,2,4  &-6.006 &-6.726 &-6.674 &-7.953 &
-8.142\\
\cite{CFK02}& & & 0,1,2 & 0,2,4 & -5.98&  &-6.73 & & -8.27 \\
FGS &&`$s$ and $p$-wave' \cite{MSA03} & 0,1 & 0,2,4 &  & &  & -7.116 & \\
\hline
FGS &CB& `$s$-wave' & 0,1,2  & 0,2,4  &-6.033 &-6.785 &-6.760&-8.079 &
-8.266\\
\cite{CFK02}& & &  &  & -6.02&  &-6.75 & & -8.19 \\
FGS & &`$s$ and $p$-wave' \cite{MSA03} & 0,1 & 0,2,4 & & & & -7.199 & \\
\hline 
FGS &$V_{\alpha\alpha}$=0& `$s$-wave' & 0,1,2  & 0,2,4  &-6.456 &-6.179 
&-5.532 &-5.884 & -5.785  
\end{tabular}
%\end{ruledtabular}
\end{center} 
\end{table}

\begin{table} 
\caption{Excitation energy in MeV of the $_{\Lambda}^{9}$Be $L=2$ 
excited (${\frac{3}{2}}^+,{\frac{5}{2}}^+$) spin-flip doublet, 
calculated for various ${\alpha\alpha}$ and $\alpha\Lambda$ potentials. 
$E_{\rm exc}^{\rm exp}(2^+)=3.04 \pm 0.03$ MeV {\protect \cite{AAC02}}.}  
\label{tab:tabl5} 
\begin{center} 
\begin{tabular}{cccccccc} 
Ref. &${\alpha\alpha}$&${\alpha\Lambda}$ & TH & Gibson & MS & MSA & 
Isle(DA) \\ 
\hline 
FGS &AB& `$s$-wave' & 2.515 & 2.593 & 2.658 & 2.847 & 2.901 \\ 
\cite{CFK02}& & & 2.73  &       & 2.76  &       & 2.92  \\ 
FGS &  & `$s$ and $p$-wave' &  &    &       & 3.160 &       \\ 
\hline
FGS &CB& `$s$-wave' & 2.804 & 2.919 & 2.956 & 3.095 & 3.144 \\ 
\cite{CFK02}& & & 2.85  &       & 2.95  &       & 3.14  \\ 
\cite{PC91} & & & 4.08  & 3.61  &       &       & 3.66  \\ 
FGS &  & `$s$ and $p$-wave' &  &    &       & 3.366 &        
\end{tabular} 
\end{center} 
\end{table}

%\newpage

%\newpage 

\begin{thebibliography}{00}

\bibitem{Tak01} H. Takahashi et al., Phys. Rev. Lett. 87 
(2001) 212502. 

\bibitem{FGa02b} I.N. Filikhin, A. Gal, Nucl. Phys. A 707 (2002) 
491. 

\bibitem{FGa02a} I.N. Filikhin, A. Gal, Phys. Rev. C 65 (2002) 
041001(R). 

\bibitem{FGS03} I.N. Filikhin, A. Gal, V.M. Suslov, Phys. Rev. C 
68 (2003) 024002. 

\bibitem{MSA03} K.S. Myint, S. Shinmura, Y. Akaishi, Eur. Phys. J.
A 16 (2003) 21. 

\bibitem{AG03} I.R. Afnan, B.F. Gibson, Phys. Rev. C 67 (2003) 
017001. 

\bibitem{LY04} D.E. Lanskoy, Y. Yamamoto, Phys. Rev. C 69 (2004) 
014303. 

\bibitem{Yam04} T. Yamada, Phys. Rev. C 69 (2004) 044301. 

\bibitem{FKM04} Y. Fujiwara, M. Kohno, K. Miyagawa, Y. Suzuki, 
J.-M. Sparenberg, arXiv: nucl-th/0405056. 

\bibitem{HKM97} E. Hiyama, M. Kamimura, T. Motoba, T. Yamada, 
Y. Yamamoto, Prog. Theor. Phys. 97 (1997) 881. 

\bibitem{MBI83} T. Motoba, H. Bando, K. Ikeda, Prog. Theor. Phys. 
70 (1983) 189.

\bibitem{PC91} O. Portilho, S.A. Coon, J. Phys. G 17 (1991) 1375. 

\bibitem{OKS00} S. Oryu, H. Kamada, H. Sekine, H. Yamashita, 
M. Nakazawa, Few-Body Syst. 28 (2000) 103. 

\bibitem{CFK02} E. Cravo, A.C. Fonseca, Y. Koike, Phys. Rev. C 66 
(2002) 014001.

\bibitem{FMK04} Y. Fujiwara, K. Miyagawa, M. Kohno, Y. Suzuki, D. Baye, 
J.-M. Sparenberg, arXiv: nucl-th/0404071. 

\bibitem{MFa93} L.D. Faddeev, S.P. Merkuriev, {\it Quantum
Scattering Theory for Several Particle Systems} (Kluwer Academic,
Dordrecht, 1993).

\bibitem{AB66} S. Ali, A.R. Bodmer, Nucl. Phys. 80 (1966) 99. 

\bibitem{FJ96} D.V. Fedorov, A.S. Jensen, Phys. Lett. B 389 
(1996) 631. 

\bibitem{CB74} W.S. Chien, R.E. Brown, Phys. Rev. C 10 (1974) 1767. 

\bibitem{WTB86} X. Wang, H. Takaki, H. Bando, Prog. Theor. Phys. 
76 (1986) 865.

\bibitem{Dav91} D.H. Davis, in: {\it LAMPF Workshop on ($\pi,K$) Physics}, 
eds. B.F. Gibson, W.R. Gibbs, M.B. Johnson, AIP Conf. Proc., Vol. 224 
(AIP, New York, 1991) pp. 38-48. 

\bibitem{The65} Y.C. Tang, R.C. Herndon, Phys. Rev. 138 (1965) B 637; 
see also R.H. Dalitz, B.W. Downs, Phys. Rev. 111 (1958) 967. 

\bibitem{GGW72} B.F. Gibson, A. Goldberg, M.S. Weiss, Phys. Rev. C 6 
(1972) 741. 

\bibitem{DGN82} C. Daskaloyannis, M. Grypeos, H. Nassena, 
Phys. Rev. C 26 (1982) 702. 

\bibitem{MS84} S. Maeda, E.W. Schmid, in: {\it Few-Body Problem in 
Physics}, ed. B. Zeitnitz (Elsevier, Amsterdam, 1984) Vol. II, 379. 

\bibitem{KAT85} Y. Kurihara, Y. Akaishi, H. Tanaka, Prog. Theor. Phys. 
71 (1984) 561; Phys. Rev. C 31 (1985) 971. 

\bibitem{ND75} M.M. Nagels, T.A. Rijken, J.J. de Swart, Phys. Rev. D 
12 (1975) 744; 15 (1977) 2547. 

\bibitem{SRi99} V.G.J. Stoks, Th.A. Rijken, Phys. Rev. C 59 
(1999) 3009. 

\bibitem{IBM85} K. Ikeda, H. Bando, T. Motoba, Prog. Theor. Phys. Suppl. 
81 (1985) 147. 

\bibitem{BSS96} J. Bernabeu, V.M. Suslov, T.A. Strizh, S.I. Vinitsky,
Hyperfine Interactions 101/102 (1996) 391. 

\bibitem{HKM02} E. Hiyama, M. Kamimura, T. Motoba, T. Yamada, 
Y. Yamamoto, Phys. Rev. C 66 (2002) 024007. 

\bibitem{CAG97} S.B. Carr, I.R. Afnan, B.F. Gibson, Nucl. Phys. A 625 
(1997) 143.

\bibitem{AAC02} H. Akikawa et al., Phys. Rev. Lett. 88 (2002) 082501. 

\bibitem{BUs87} A.R. Bodmer, Q.N. Usmani, Nucl. Phys. A 468 (1987) 653. 
 
\end{thebibliography}
\end{document}